\documentstyle[12pt]{article}
\begin{document}

\baselineskip 20pt 

\vspace{.5in}

\begin{center}

{\Large \bf   
Two types of electronic states\\ in \\
one dimensional crystals of finite length
}
\vskip .8in
{\bf
Shang Yuan Ren\\
Department of Physics, Peking University \\
Beijing 100871, People's Republic of China\\
}
\vbox{}
\end{center}
\newpage
\begin{center}
{\bf \large Abstract}
\end{center}
\par
Exact and general results on the electronic states in one dimensional crystals
bounded at $\tau$ and $\tau + L $ - where $ L =Na$, $N$ is a positive 
integer and $a$ is the potential period  - are presented. 
Corresponding to each energy band of the Bloch wave, there are $N-1$ states 
in the finite crystal and their energies are dependent on the crystal length
$L$ but not on the crystal boundary $\tau$ and map the energy band 
exactly; There is always one and only one electronic state corresponding to
each band gap of the Bloch wave,
whose energy is dependent on the crystal boundary location $\tau$ but
not on the crystal length $L$. 
This state is either a constant energy confined band edge state or a surface
state in the band gap. A slight change of the boundary location $\tau$
could change the property and the energy of this state dramatically.
\newpage
Bloch theorem plays a central role in our current understanding
on the electronic structures in modern solid state physics.
However, any real crystal always has a finite size and does not have a 
hypothetical infinite size or the
periodic boundaries which Bloch theorem is based on\cite{cal}. 
The difference between the electronic structure of a real crystal of 
finite size and the electronic structure 
obtained based on the translational invariance becomes more significant
as the crystal size decreases. A clear understanding of the properties
of electronic states in real crystals of finite size has 
both theoretical and practical significant importance. A straightforward 
way is to obtain exact solutions for crystals of finite size.
However, to obtain exact solutions of the Schr$\ddot{\rm o}$dinger 
equation with a general periodic potential for a finite crystal with 
boundaries has been being considered as a rather difficult problem: the 
lack of translational invariance in the crystal of finite size is a 
major obstacle. It is the use of the
translation invariance that greatly simplifies the mathematics in
solving the Schr$\ddot{\rm o}$dinger equation with a periodic potential.
Without a such simplification
the corresponding problem for finite crystals with boundaries 
could become rather difficult mathematically. 
Thus most previous theoretical investigations on this subject were based on 
approximate and/or numerical approaches and were usually on a specific 
material and/or on a specific model\cite{rvs}.\par
Historically, many of our current fundamental understandings on the 
electronic structures of crystals were obtained through the analysis of one 
dimensional crystals\cite{kit,sei,jon}. Among the most well known examples are 
the Kronig-Penney model\cite{kro}, Kramers' analysis of on the band 
structure of one dimensional infinite crystals\cite{kra}, the Tamm's
surface state\cite{tam}, etc.  Kramers' analysis
on the solutions of the one dimensional Schr$\ddot{\rm o}$dinger equation
with periodic potential - based on a differential equation theory approach -
provided a thorough and general understanding on the band structure of
one dimensional crystals\cite{sei,jon,koh}. \par
In this work we present exact and general results on the electronic states in 
one dimensional crystals of finite length $ L = Na$ 
- where $a$ is the potential period and $N$ is a positive integer - 
in comparison with the band structure of one dimensional infinite crystal,
using a differential equation theory approach\cite{eas}. It also shows
that the obstacle or mathematical difficulty due to the lack of translation
invariance in fact can be circumvented.
Different from Kramers' work, the understanding on the $zeros$ of solutions
of one dimensional Schr$\ddot{\rm o}$dinger differential equation with 
periodic potential\cite{eas} plays a fundamental role in this work. 
\par
The one dimensional Schr$\ddot{\rm o}$dinger differential equation with
a periodic potential can be written as 
\begin{equation}
- y'' (x) +[ v(x) - \lambda ] y (x) = 0,
~~~~~~~~~~~~~~~~~~~~~~~~~~~~-\infty~<~x~<~+\infty
\end{equation}
where
$$
v(x+a) = v(x).
~~~~~~~~~~~~~~~~~~~~~~~~~~~~~~~~~~~~~~~~~~~~~~~~~
$$
We assume (1) is solved, all solutions are known. The eigenvalues are
energy bands $\varepsilon_n(k)$ and the corresponding eigenfunctions are
Bloch functions $\phi_n(k,x)$, where $n~=~0,~1,$ $~2,~.....$. 
\par
For a one dimensional crystal with two boundaries at $\tau$ and $\tau + L$, 
we assume the potential $inside$ the
crystal is still $v(x)$ as in (1) and all electronic states in the crystal are
confined in the finite size of the crystal.
We are looking for the eigenvalues $\Lambda$ and eigenfunctions
$\psi(x, \Lambda)$, which are solutions of
\begin{equation}
- y'' (x) +[ v(x) - \Lambda ] y(x) = 0,
~~~~~~~~~\tau~<~x~<~\tau + L
\end{equation}
and
\begin{equation}
~~~~~~~~~ y( x) = 0,~~~~~~~~~~~~~~~~~~~~~~~~~~~~~~~~~~
 x~\leq~\tau~~or~~ x~\geq~\tau +L
\end{equation}
where $\tau$ is a real number, $ L = Na$ and $N$ is a positive integer.
\par
By using a differential equation approach we found that 
all solutions of (2) and (3) can be obtained if all solutions of (1) are known
(See Appendix A) and the electronic states in one-dimensional finite crystals 
can be classified into two types:
\\
1. Inside an energy band, there are $N-1$ electronic states whose energy is
dependent on the crystal length $L$ and given by:
\begin{equation}
k(\Lambda_j) = j~\pi/L, ~~~~~~~~~~~~j = 1, 2, ....N-1. 
\end{equation}
Correspondingly, there are $N-1$ eigenfunctions $\psi(x, \Lambda_j)$.
Each eigenvalue for this case is a function of $L$, the crystal length.
But they all do not depend on the location of the crystal boundary 
$\tau$ or $\tau~+~L$. For simplicity, we call these
states as $L$-dependent states, although only the eigenvalue
of a such state is dependent on only $L$, the wavefunction
of a such state is dependent on both $\tau$ and $L$.
\par
The eigenvalues $k(\Lambda_j)$ map the dispersion relation 
$k(\varepsilon_n)$ of the original equation (1) exactly.
Many authors noted that there is an approximate 
correspondence between the bulk energy dispersion and the 
energy levels in crystals of finite size\cite{zz,ped,pop}, from Eq. (4) 
we see that this
is in fact an exact correspondence for the electronic states in general one 
dimensional crystals of finite length. Furthermore this exact 
correspondence does not depend on the crystal boundary $\tau$.
Pedersen and Hemmer\cite{ped} investigated the non-bandedge electronic states 
in one dimensional finite crystals 
with a Kronig-Penney model and also found that
the energy spectrum is independent on the crystal boundary. 
\\
2. For each band gap, there is always one and only one electronic states,
whose energy $\Lambda$ is given by a necessary and sufficient condition:
\begin{equation}
\psi(\tau~+~a, \Lambda)~ =~\psi(\tau, \Lambda)~=~0. 
\end{equation}
\par
(5) does not contain the crystal length $L$, thus the energy $\Lambda$
corresponding to a band gap is dependent on the boundary of the 
crystal $\tau$, but not on the crystal length.
For simplicity, we call these
solutions as $\tau$-dependent states, although only the eigenvalue
of a such state is dependent on only $\tau$, the wavefunction
of a such state is dependent on both $\tau$ and $L$.
The energy $\Lambda$ of these $\tau$-dependent states
can be labeled as $\Lambda_{\tau, 2m}$ or $\Lambda_{\tau, 2m+1}$,
corresponding to the band gap at $k = \pi/a$ or $k =0$. 
$\Lambda_{\tau, 2m}$ is in 
$[\varepsilon_{2m}(\pi/a), \varepsilon_{2m+1}(\pi/a)]$ and 
$\Lambda_{\tau, 2m+1}$ is in 
$[\varepsilon_{2m+1}(0), \varepsilon_{2m+2}(0)]$. 
\par 
How these $\tau$ dependent eigenvalues change
as the crystal boundary $\tau$ changes can be seen in Fig. 1,
using $\Lambda_{\tau, 1}$ as an example. The details can be found
in Appendix B. In Fig. 1 the zeros of band edge wavefunctions 
$\phi_1(0,x)$ and $\phi_2(0,x)$ are shown as circles.
\par
If $\tau$ is equal to a zero of a band edge wavefunction $\phi_1(0,x)$ or 
$\phi_2(0,x)$, $\Lambda_{\tau, 1}$ is equal to $\varepsilon_1(0)$
or $\varepsilon_2(0)$.
In any one of these cases, we have an electronic state whose energy 
is the corresponding band edge energy of (1),
independent on the crystal length $L$.
Zhang and Zunger first observed a state with such behavior in their
numerical investigations on Si quantum films\cite{zz}. 
Franceschetti and Zunger also observed a such state in their 
calculations on the free standing GaAs quantum film\cite{fz}. 
The author has pointed out that the existence of the band edge states
with such behavior in fact could be quite general
in one dimensional symmetric finite crystals\cite{syr}.
From this work we understand that for the electronic states in one 
dimensional finite crystals, without any requirement on the symmetry
of the periodic potential or of the system, {\it as long as a 
boundary is a zero of a band edge wavefunction of (1), there always is a 
confined state
whose energy is the band edge energy, independent on the crystal length $L$}. 
Due to the inversion symmetry of the finite crystals treated in \cite{syr},
for each band gap there is always one band edge wavefunction which
is zero at the lattice boundaries.
\par
If $\tau$ is not a zero of either one of the two band edge wavefunctions, 
i.e, $\tau$ is in the dotted line region or in the dashed line region in 
Fig. 1, then the electronic state $\psi(x, \Lambda)$ will have the form of 
either $e^{\beta(\Lambda) x} p_1(x, \Lambda)$ or 
$e^{- \beta(\Lambda) x} p_2(x, \Lambda)$ with an 
energy $\Lambda$ $inside$ the band gap. This is a surface state.
\par
Many years ago, by using a Kronig-Penney model Tamm\cite{tam} showed that 
the termination of the periodic potential at the surface of a crystal 
can cause a surface state existing in the band gap for each band gap of the 
Bloch wave. More than sixty years later, Zhang and Zunger\cite{zz} and 
Franceschetti and Zunger\cite{fz} observed the existence of the constant 
energy confined band edge state in their numerical calculations.
Now we understand that in the one dimensional case,
the surface state in the gap and the constant
energy confined state at a band edge are two different results
of the termination of the periodic potential, depending on
whether the boundary $\tau$ is a zero of a band edge wavefunction.
It may be noticed that a
Tamm's surface state is an extra "added" state, the $\tau$-dependent states
in this work comes from a unified solving of the all states.\par
An interesting question is: Why a finite one dimensional crystal of 
$two~ends$ can only have $at~most$ $one$ surface state in each gap? 
This is due to that the two ends of a finite crystal in general
are not equivalent, except in a $symmetric$ finite crystal\cite{neq}.
A symmetric one dimensional finite crystal requires:
(1) The crystal potential has an inversion symmetry center. 
(2) The two ends of the crystal are symmetric to the same 
symmetry center and thus are equivalent. If any one of these two is not 
true, then the two ends of the finite crystal are in fact not 
equivalent. For a symmetric finite one dimensional crystal, it has one 
constant energy confined band edge state, rather than a surface state, in 
each gap. That is the result obtained in \cite{syr}.
\par
A little further thinking can also help to understand this unexpected 
result. For simplicity we assume a monoatomic crystal of $L=Na$ here.
If there are $N-1$ bulk states in each
band, and there are altogether $N$ atoms, there is only one, not two, more
state which can be the either the surface state or at the band edge. In the 
case of $N=1$, there is one state for each energy band. 
Similarly, one monolayer of atoms also can have only one two-dimensional 
energy band for each atomic state, even it has two surfaces. 
The author has also worked out a similar problem on three dimensional finite 
crystals\cite{syr2}. For a finite crystal of parallelogram shape which has 
$N_1{\bf a}_1$, $N_2{\bf a}_2$ and $N_3{\bf a}_3$ forming the sides which 
meet at a corner, 
for each energy band there are $(N_1-1)(N_2-1)(N_3-1)$ bulk states,
$(N_1-1)(N_2-1) + (N_2-1)(N_3-1) + (N_3-1)(N_1-1)$ surface states,
$(N_1-1) + (N_2-1) + (N_3-1)$ "side states" and one "conner state" in spite
of that the crystal has six faces, twelve sides and eight corners:
The sum of these numbers is exact $N_1 N_2 N_3$. Any change of these numbers
will not give the correct sum. 
\par
A slight change of the boundary 
location $\tau$ can change the properties of the $\tau$ dependent 
state dramatically, this can be
clearly seen in Fig. 1: If $\tau$ is in the region corresponding to a dotted
line, the surface state is near one end of the crystal.
If $\tau$ is in the region corresponding to a dashed
line, the surface state is near the other end of the crystal.
If $\tau$ is a zero of a band edge wavefunction
(either a solid circle or an open circle), then $\tau + L$
is also a zero of the same band edge wavefunction and thus the boundary $\tau$
dependent state is a constant energy confined band edge state.
\par
As an example, in Fig. 2 is shown a comparison between the the energy bands 
and the eigenvalues of the electronic states 
in an one dimensional crystal of length $L = 8a$.
In Fig. 3. are shown the energies of three electronic states in crystals
of finite length near or
in the lowest band gap of the Bloch wave as functions of the crystal 
length $L$. 
\par
In summary, by using a differential equation theory approach, we have  
obtained exact and general results on all electronic states
in one dimensional crystals of finite length in comparison with the
electronic states in a periodic potential obtained from (1).
For one dimensional crystals bounded at $\tau$ and $\tau + L$, 
there are two different types of electronic states: There are $N-1$ states 
corresponding to each energy band of (1).
Their eigenvalues $\Lambda$ are given by (4), thus are dependent on the crystal
length $L$ but not on the crystal boundary location $\tau$
and map the energy band exactly; 
There is always one and only one electronic state corresponding to
each band gap of (1),
whose eigenvalue $\Lambda$ is dependent on the boundary location $\tau$ but
not on the crystal length $L$. 
Such a $\tau$-dependent state can be either a constant energy confined
band edge state (if the crystal boundary $\tau$ is a zero of a 
band edge wavefunction of (1)) or a surface state in the band gap 
(if the crystal boundary $\tau$ is not a zero of either band 
edge wavefunction of (1)). A slight change of the crystal boundary location
could change the properties and the energy of this boundary dependent state
dramatically\cite{hi}.
\par
The results obtained in \cite{syr} are a special case of the more general
results obtained here.
\vskip .2in
Acknowledgment:
\par
The author is grateful to Professors Kun Huang, Lo Yang, W. A. Harrison
and Peter Y. Yu and Dr. L. W. Wang for
stimulating discussions. This research is supported by the National Natural
Science Foundation of China (Project No. 19774001). 

\newpage
\begin{center} 
Appendix A
\end{center}
The Schr$\ddot{\rm o}$dinger differential equation (1) can be considered as a 
special case of the more general Hill's equation\cite{eas}.
This work is mainly based on the general properties of solutions
of Hill's equation.\par
We are mainly interested in the cases for which there is always a gap
between two consecutive energy bands of (1)\cite{dbeg}. For these Bloch waves,
the band edges $\varepsilon_n(0)$ and $\varepsilon_n(\pi/a)$ occur in the 
order
$$
\varepsilon_0(0) < \varepsilon_0(\pi/a) < \epsilon_1(\pi/a) < 
\varepsilon_1(0) < 
\varepsilon_2(0) < \varepsilon_2 (\pi/a)< \varepsilon_3(\pi/a)
< \varepsilon_3(0)  < \varepsilon_4(0) < .....
$$
The band gaps are between $ \varepsilon_{2m} (\pi/a)$ and 
$ \varepsilon_{2m+1}(\pi/a)$
or between $ \varepsilon_{2m+1} (0)$ and $< \varepsilon_{2m+2}(0)$.
\par
\par
Suppose $y_1(x, \lambda)$ and $y_2(x, \lambda)$ are two linearly independent
solutions of (1), in general, a solution of (2) and
(3) if it exists, can be expressed as
$$
\psi( x, \Lambda) = f( x, \Lambda) ,~~~~~~~ if~~\tau~<~x~<~\tau +L
$$
$$
~~~~~~~~~~~~~~~~~ = 0,~~~~~~~~~~~~~~ if~~x~\leq~\tau~~or~~ x~\geq~\tau + L
$$
where
$$
f( x, \lambda) =  c_1 y_1(x, \lambda) + c_2 y_2(x, \lambda)
\eqno (A.1)
$$
is a non-trivial solution of (1) and satisfies
$$
f( \tau, \Lambda) = f( \tau + L, \Lambda) = 0.
\eqno (A.2)
$$
The non-trivial solutions of (2) and (3) can be found through (A.2) 
based on the general properties of solutions of (1). \par
The properties of linearly independent solutions of (1) are determined by
a real number $D(\lambda)$ called its discriminant\cite{eas}: 
Suppose $\eta_1(x, \lambda)$ and $\eta_2(x,\lambda)$ are two linearly 
independent solutions of (1), which satisfy the initial conditions
$
\eta_1(0, \lambda) = 1,~~
\eta_1'(0, \lambda) = 0;~~~
\eta_2(0, \lambda) = 0,~~
\eta_2'(0, \lambda) = 1.
$
The real number $D(\lambda)$ defined by
$
D (\lambda) = \eta_1(a, \lambda) + \eta_2'(a, \lambda)
$
is called the discriminant of (1). 
\par
As $\lambda$ increases from $-\infty$ to $+\infty$, the value of $D(\lambda)$
changes as the following (m = 0, 1, 2, ....)\cite{eas,kra,koh}:
\par
(i) In the intervals $[\varepsilon_{2m}(0), \varepsilon_{2m}(\pi/a)]$, 
$D(\lambda)$ decreases from 2 to -2.
\par
(ii) In the intervals $[\varepsilon_{2m+1}(\pi/a), \varepsilon_{2m+1}(0)]$, 
$D(\lambda)$ increases from -2 to 2.
\par
(iii) In the intervals $(-\infty, \varepsilon_0(0))$ and 
$(\varepsilon_{2m+1}(0), \varepsilon_{2m+2}(0))$, $D(\lambda) > 2$.
\par
(iv) In the intervals $(\varepsilon_{2m}(\pi/a), \varepsilon_{2m+1}(\pi/a))$, 
$D(\lambda) < -2$.
\par
The existence of and the properties of non-trivial solutions $\Lambda$ and 
$f(x, \Lambda)$ in (A.2) can be determined on this basis. 
\par
For solutions of (2) and (3), both the permitted and the forbidden 
eigenvalue ranges of (1) cannot be excluded. We should consider
solutions of (A.2) for $\lambda$ in $(-\infty, +\infty)$.
However, according to the Theorem 3.2.2 of \cite{eas}, there is not
a nontrivial solution of (A.2) for which 
$\Lambda$ in $(- \infty, \varepsilon_0(0)]$.
Thus we need only to consider $\lambda$ in $(\varepsilon_0(0), +\infty)$.
Depending on $\lambda$, the value of $D(\lambda)$ may have five different 
cases\cite{eas}: 
\\
Case A. $|D(\lambda)| < 2$.
\par
In this case $\lambda$ is inside an energy band of (1).
Two linearly independent solutions of (1) can be expressed as\cite{eas}
$$
y_1(x, \lambda) = e^{i k(\lambda) x} p_1(x, \lambda),~~
y_2(x, \lambda) = e^{- i k(\lambda) x} p_2(x, \lambda),
$$
where $k(\lambda)$ is a real number depending on $\lambda$ and
\begin{eqnarray*}
0 < k(\lambda) a < \pi 
\end{eqnarray*}
and $p_1(x, \lambda)$ and $p_2(x, \lambda)$ have period $a$:
$ p_i(x +a, \lambda) = p_i(x, \lambda)$.
All $k(\lambda)$ and $p_i(x, \lambda)$ are functions
of $\lambda$.
\par
If there is a non-trivial solution $f(x, \Lambda )$ of (A.2), simple
mathematics gives that we must have either 
$$
e^{i k(\Lambda) L} - e^{- i k(\Lambda) L} = 0 ,
\eqno(A.A.1)
$$ 
or
$$
c_1 p_1(\tau, \Lambda) = 0, ~~and~~ c_2 p_2( \tau, \Lambda) = 0.
\eqno(A.A.2)
$$
We can easily prove that neither $p_1(\tau, \Lambda)$ nor 
$p_2(\tau, \Lambda)$ can be zero. Suppose $p_1(\tau, \Lambda) = 0$, 
we must have $p_1(\tau +a, \Lambda) = 0$
and thus $y_1(\tau, \Lambda) = y_1(\tau +a, \Lambda) = 0$. Then 
according to theorem 3.1.3 of \cite{eas}, $\Lambda$ must be in  
$[\varepsilon_{2m}(\pi/a), \varepsilon_{2m+1}(\pi/a)]$
or in $[\varepsilon_{2m+1}(0), \varepsilon_{2m+2}(0)]$, in which 
$ | D(\Lambda) | \geq 2$. This is in
contradictory to $ | D(\lambda)| < 2 $ for the Case A. Similarly 
$p_2(\tau, \Lambda)$ can not be zero. Thus 
no non-trivial solution obtained from $(A.A.2)$ exists.
\par 
Therefore non-trivial solutions in this case can only be obtained by $(A.A.1)$.
Note $(A.A.1)$ does not contain $\tau$.
The non-trivial solutions can be obtained if 
$$
k(\Lambda) L = j \pi,~~~~~~~~~~~~ 
$$
Thus in each energy band $k(\varepsilon_n)$, there are $N-1$ values
of $\Lambda_j$, where $j = 1, 2, ....N-1$, for which
$$
k(\Lambda_j) = j~\pi/L.
$$
This is Eq. (4) in the text.
\\
Case B. $D(\lambda) = 2$.
\par
In this case $\lambda$ is at a band edge at $k~=~0$: 
$\lambda~=~ \varepsilon_{2m+1}(0) $ or $\lambda~=~ \varepsilon_{2m+2}(0) $. 
\par
Two linearly independent solutions of (1) can be expressed as\cite{eas}
$$
y_1(x, \lambda) = p_1(x, \lambda),~~
y_2(x, \lambda) = x p_1(x, \lambda) + p_2(x, \lambda),
$$
and $p_1(x, \lambda)$ and $p_2(x, \lambda)$ have period $a$.
\par
Due to the Sturm Separation Theorem\cite{eas2}, the zeros of 
$p_1(x, \lambda) $ are separated from the zeros of $ p_2(x, \lambda) $. 
Simple mathematics leads to that
the existence of non-trivial solution (A.2) in this case requires
$$
p_1( \tau, \Lambda) = 0  
~~and~~
c_2 = 0.
\eqno (A.B.1)
$$
\\
Case C. $ D(\lambda) > 2$.
\par
In this case $\lambda$ is $inside$ a band gap 
at $k~=~0$: 
$\varepsilon_{2m+1}(0)~<~ \lambda~ <~ \varepsilon_{2m+2}(0)$.
Two linearly independent solutions of (1) can be expressed as\cite{eas}
$$
y_1(x, \lambda) = e^{\beta(\lambda) x} p_1(x, \lambda),~~
y_2(x, \lambda) = e^{-\beta(\lambda) x} p_2(x, \lambda),
$$
where $\beta(\lambda)$ is a positive real number depending on $\lambda$ and
$p_1(x, \lambda)$ and $p_2(x, \lambda)$ have period $a$.
\par
Again due to the Sturm Separation Theorem\cite{eas2}, the zeros of 
$p_1(x, \lambda) $ are separated from the zeros of $ p_2(x, \lambda) $. 
If there is a non-trivial solution $f(x, \Lambda)$ in this case, 
simple mathematics from (A.1) and (A.2) gives that we must have either
$$
p_1(\tau, \Lambda) = 0,~~and~~~c_2~=~0 
\eqno(A.C.1)
$$
or
$$
p_2(\tau, \Lambda) = 0,~~and~~~c_1~=~0.
\eqno(A.C.2)
$$
Note that $(A.C.1)$ and $(A.C.2)$ cannot be true simultaneously.
\par
From the discussion on Case B and Case C, we can see that
if we have a non-trivial solution of (A.2) for a band gap at $k=0$ 
($\varepsilon_{2m+1}(0)~\leq~ \lambda~ \leq~ \varepsilon_{2m+2}(0)$),
either one of (A.B.1), (A.C.1) or (A.C.2) must be true.
Since all functions $p_i(x, \lambda)$ in (A.B.1), (A.C.1) and (A.C.2) are 
periodic functions, we always have $f(\tau~+~a, \Lambda) = 0$ if we 
have $f(\tau, \Lambda) =0$. Therefore the 
following equation is a necessary condition for having a solution
$\Lambda$ in (A.2) for a band gap at $k=0$:
$$
f(\tau~+~a, \Lambda)~ =~f(\tau, \Lambda)~=~0. 
\eqno (A.3)
$$
It is easy to see that (A.3) is also a sufficient condition for having
a solution (A.2): From (A.3) one can obtain
$f(\tau~+~n~a, \Lambda)~ = ~0$, where $n = 0,...N$. 
\par
Corresponding to a band gap at $k=\pi/a$,
the Case D ($D(\lambda)= -2)$ and the Case E ($D(\lambda) < - 2$) can be
very similarly discussed as the Case B and the Case C. The major difference
is there we have semi-periodic functions 
($s_i(x + a, \Lambda) = - s_i(x, \Lambda)$) instead of periodic
functions. We are led to the same equation (A.3) as a necessary and
sufficient condition for having a solution (A.2) for a band gap at $k=\pi/a$.
Thus (A.3) is a necessary and sufficient condition
for having a solution (A.2) corresponding to a band gap.
(A.3) is essentially the same as Eq. (5) in the text.
\par
The Theorem 3.1.3 of \cite{eas} indicates that for an arbitrary real number
$\tau$, there is always one and only one $\Lambda$ for which (A.3) is true 
for each band gap
$[\varepsilon_{2m}(\pi/a), \varepsilon_{2m+1}(\pi/a)]$ or 
$[\varepsilon_{2m+1}(0), \varepsilon_{2m+2}(0)]$. 
\par
Therefore {\it for any real number $\tau$ there is always one and 
only one $\Lambda$,
which is a solution of (A.2) and dependent on $\tau$ but not $L$, for each 
band gap $[\varepsilon_{2m}(\pi/a), \varepsilon_{2m+1}(\pi/a)]$ or 
$[\varepsilon_{2m+1}(0), \varepsilon_{2m+2}(0)]$}. 
\newpage
\begin{center}
Appendix B
\end{center}
\par
How these $\tau$-dependent eigenvalues $\Lambda$ change as $\tau$ changes
can be obtained from differential equation theory.  Here we 
discuss $\Lambda_{\tau, 2m+1}$ as an example. $\Lambda_{\tau, 2m}$ can be very
similarly discussed.
\par 
According to the theorem 3.1.2 of \cite{eas}, $\phi_{2m+1}(0,x)$ and 
$\phi_{2m+2}(0,x)$ have exact $2m+2$ zeros in $[0,a)$. 
Then according to the Sturm Comparison Theorem\cite{eas2}, the zeros of 
$\phi_{2m+1}(0,x)$ and $\phi_{2m+2}(0,x)$
must be distributed alternatively:
There is always one and only one zero of $\phi_{2m+2}(0,x)$ between
two consecutive zeros of $\phi_{2m+1}(0,x)$, and
there is always one and only one zero of $\phi_{2m+1}(0,x)$ between
two consecutive zeros of $\phi_{2m+2}(0,x)$.
\par
We treat $\tau$ as a variable and let $\tau$ go continuously 
from a (any) zero $x_{1,2m+1}$ of $\phi_{2m+1}(0,x)$ to
$x_{1,2m+2}$, the zero of $\phi_{2m+2}(0,x)$ that is next to $x_{1,2m+1}$. 
Since $\Lambda_{\tau, 2m+1}$ is a continuous function of $\tau$\cite{eas},
the corresponding $\Lambda_{\tau, 2m+1}$ will 
also go continuously from $\varepsilon_{2m+1}(0)$ to $\varepsilon_{2m+2}(0)$. 
Similarly if $\tau$ goes continuously from $x_{1,2m+2}$ to $x_{2,2m+1}$,
the next zero 
of $\phi_{2m+1}(0,x)$, the corresponding $\Lambda_{\tau, 2m+1}$ 
will also go back continuously from 
$\varepsilon_{2m+2}(0)$ to $\varepsilon_{2m+1}(0)$. 
In Figure 1 is shown $\Lambda_{\tau, 1}$ as function of $\tau$
in the interval $[x_{1,1}, x_{1,1} + a]$, where $x_{1,1}$ is a zero
of $\phi_{1}(0,x)$.  
\par
If we consider as $\tau$ goes from $x_{1,2m+1}$ to $x_{1,2m+2}$ and then
to $x_{2,2m+1}$, the corresponding $\Lambda_{\tau, 2m+1}$ 
as function of $\tau$
goes up from $\varepsilon_{2m+1}(0)$ to $\varepsilon_{2m+2}(0)$ 
and then back to $\varepsilon_{2m+1}(0)$ 
as a basic undulation, in this basic undulation the function 
$f(x, \Lambda)$ has different forms. 
We know for any solution (A.2) in each band gap 
$[\varepsilon_{2m+1}(0), \varepsilon_{2m+2}(0)]$,
either one of (A.B.1), (A.C.1) or (A.C.2) must be true. 
When $\Lambda = \varepsilon_{2m+1}(0)$ or 
$\Lambda = \varepsilon_{2m+2}(0)$, (A.B.1) is 
true. For the two sections (end points excluded) of this basic undulation
either (A.C.1) or (A.C.2) is true.
In one section $f(x, \Lambda)$ has the form 
$ c_1 e^{\beta(\Lambda)~x}~p_1(x, \Lambda)$ (A.C.1 is true)
and in the other section $f(x, \Lambda)$ 
has the form $ c_2 e^{- \beta(\Lambda)~x}~p_2(x, \Lambda)$ (A.C.2 is true).
In which section it has which form is dependent on $v(x)$ in (1). In Fig. 1 
the two sections of a basic undulation are shown as a dashed line 
and a dotted line, indicating two different forms of
$f(x, \Lambda)$. Since in the interval [0, a), 
both $\phi_{2m+1}(0,x)$ and $\phi_{2m+2}(0,x)$ have exactly $2m+2$ zeros,
then in general $\Lambda_{\tau, 2m+1}$ as function of $\tau$ will
always complete $2m+2$ basic undulations in an interval of length $a$.
Similarly $\Lambda_{\tau, 2m}$ as function of $\tau$ will
always complete $2m+1$ basic undulations in an interval of length $a$.
\par
A function with the form of $ c_1 e^{\beta(\Lambda)~x}~p_1(x, \Lambda)$ 
or $ c_2 e^{- \beta(\Lambda)~x}~p_2(x, \Lambda)$, in which 
$\beta(\Lambda) > 0$,
is mainly distributed near either one of the two boundaries
of the finite crystal,
due to the exponential factor. Thus these $\tau$-dependent 
states are in fact the surface states introduced by the termination of the 
periodic potential.  
{\it These surface states are introduced into the band gaps when 
the boundary $\tau$ is 
not a zero of either band edge wavefunction of the original Bloch waves}. 
Of course, for the band gaps at $k=\pi/a$
($\varepsilon_{2m}(\pi/a) < \Lambda < \varepsilon_{2m+1}(\pi/a)$),
a surface state has
the form of either $ c_1 e^{\beta(\Lambda)~x}~s_1(x, \Lambda)$ 
or $ c_2 e^{- \beta(\Lambda)~x}~s_2(x, \Lambda)$ inside the crystal,
where $s_1(x, \Lambda)$ and $s_2(x, \Lambda)$ are semi-period functions.

\newpage
\begin{center} 
Figure Captions
\end{center}

Fig. 1. $\Lambda_{\tau, 1}$ as function of $\tau$ 
in the interval $[x_{1,1}, x_{1,1} + a]$. 
The zeros of $\phi_1(0,x)$ are shown as solid circles and the
zeros of $\phi_2(0,x)$ are shown as open circles. Note that
 $\Lambda_{\tau, 1}$ completes two basic undulations in  
$[x_{1,1}, x_{1,1} + a]$. The dashed lines and dotted lines can be considered
as the surface state is located near two different ends of the finite crystal.\\
   \\

Fig. 2. A comparison between the energy bands $\varepsilon_n(k)$ of (1) 
(solid lines) and the eigenvalues $\Lambda$ of the electronic states in a 
crystal of length $~=~8a$
(solid circles, $L$-dependent; open circles, $\tau$-dependent).
Note that: the $L$-dependent
eigenvalues map the energy bands exactly and satisfy (4); the 
$\tau$-dependent eigenvalues are in a band gap or at a band edge of (1).\\ 
   \\

Fig. 3. The eigenvalues of three electronic states in
finite crystals near or in the lowest 
band gap of (1) as functions of the crystal length $L$,
while $\tau$ is fixed. Note the eigenvalue of the
$\tau$-dependent electronic state (open circles) in the gap is independent
on $L$, the eigenvalues of the two $L$-dependent electronic states
(solid circles) change as $L$ changes.
   \\


\newpage

\begin{thebibliography}{99}

\bibitem{cal}
See for example, J. Callaway, 
{\it  Quantum theory of the solid state}, Second edition,
Academic Press, London (1991).

\bibitem{rvs}
See for example, 
A. D. Yoffe, Adv. Phys. {\bf 42}, 173 (1993);
A. D. Yoffe, Adv. Phys. {\bf 50}, 1 (2001);
M. Kelly, {\it Low Dimensional Semiconductors: Materials, Physics, Devices
and Applications}, Oxford University Press, Oxford (1996);
Peter Y. Yu and Manuel Cardona,
{\it Fundamentals of Semiconductors, Physics and Materials Properties},
Springer Verlag, Berlin (1999).

\bibitem{kit} 
C. Kittel, {\it Introduction to Solid State Physics}, Seventh edition, 
John Wiley \& Sons, New York (1996).

\bibitem{sei} F. Seitz, {\it The Modern Theory of Solids},
McGraw-Hill, New York (1940).

\bibitem{jon} H. Jones, {\it The Theory of Brillouin Zones and Electronic
States in Crystals},
North-Holland, Amsterdam (1960).

\bibitem{kro} 
 R. L. Kronig and W. G. Penney, Proc. Roy. Soc. London. Ser. A. {\bf 130},
 499 (1931). 

\bibitem{kra} The nature of the solutions of Eq. (1) was first 
 investigated by Kramers. See, von H. A. Kramers, Physica, {\bf 2}, 
 483 (1935).
 In his paper, Kramers showed $f(E)$, which is the same as $D(\lambda)$
 in this work, as function of $E$ in a figure. See also \cite{koh}.

\bibitem{tam} I. Tamm, Physik. Z. Sowj., {\bf 1}, 733 (1932).

\bibitem{koh} W. Kohn, Phys. Rev. {\bf 115}, 809 (1959).

\bibitem{eas} M. S. P. Eastham, {\it The Spectral Theory of Periodic
Differential Equations}, Scottish Academic Press,
Edinburgh (1973) and references therein. 

\bibitem{zz} 
S. B. Zhang and A. Zunger, Appl. Phys. Lett. {\bf 63}, 1399 (1993).

\bibitem{ped} 
F. B. Pedersen and P. C. Hemmer, Phys. Rev. {\bf B50}, 7724 (1994).

\bibitem{pop} For example, Z. V. Popovic, H. J. Trodahl, M. Cardona, 
E. Richter, D. Strauch and K. Ploog, Phys. Rev. {\bf B40}, 1202 (1989);
Z. V. Popovic, M. Cardona, E. Richter, D. Strauchi, L. Tapfer and K. Ploog,
Phys. Rev. {\bf B40}, 1207 (1989);
Z. V. Popovic, M. Cardona, E. Richter, D. Strauchi, L. Tapfer and K. Ploog,
Phys. Rev. {\bf B40}, 3040 (1989);
P. Malinasmata and M. Cardona, 
Superlattices and Microstructures {\bf 10}, 39 (1991).

\bibitem{fz} 
A. Franceschetti and A. Zunger, Appl. Phys. Lett. {\bf 68}, 3455 (1996).

\bibitem{syr} 
S. Y. Ren, Phys. Rev. {\bf B64}, 035322, (2001).

\bibitem{neq} 
For a specific band gap, if $in~the~finite~crystal$ $\tau$ is closer to a 
zero of the higher band edge wavefunction, then $\tau + L$ will be 
closer to a zero of the lower band edge wavefunction, and 
vice versa. 
Only if $\tau$ $is$ a zero of a 
band edge wavefunction, then $\tau + L$ will also be a zero of the same band
edge wavefunction. If this happens, we will have a constant energy 
confined band edge state rather than a surface state for $this$ band gap.

\bibitem{syr2} 
S. Y. Ren, unpublished.

\bibitem{hi} 
By using a nearest neighbor tight binding model, Hatsugai and Iguchi found
that in one dimensional chains of finite length corresponding to each band
gap there is only $one$ confined state. 
See, Y. Hatsugai, Phys. Rev. {\bf B 48}, 11851 
(1993); K. Iguchi, Int. J. Mod. Phys. {\bf B 11}, 2157 (1997).

\bibitem{dbeg}  For the more general cases,
we may have $\varepsilon_{2m}(\pi/a)
~=~\varepsilon_{2m+1}(\pi/a)$
or $\varepsilon_{2m+1}(0) =  \varepsilon_{2m+2}(0)$. In those cases, 
it is easy to prove that there always is a solution $\Lambda$ in (5)
which is dependent on neither $L$ nor $\tau$: 
$\Lambda = \varepsilon_{2m}(\pi/a)$ or
$\Lambda =  \varepsilon_{2m+1}(0)$. $f(x, \Lambda)$ will be either a
semi-periodic function 
(when $\varepsilon_{2m}(\pi/a)~=~ \varepsilon_{2m+1}(\pi/a)$)
or a periodic function 
(when $\varepsilon_{2m+1}(0)~=~\varepsilon_{2m+2}(0)$).

\bibitem{eas2} See, for example, M. S. P. Eastham, {\it Theory of ordinary
differential equations}, Van Nostrand Reinhold, (1970). 

\end{thebibliography}
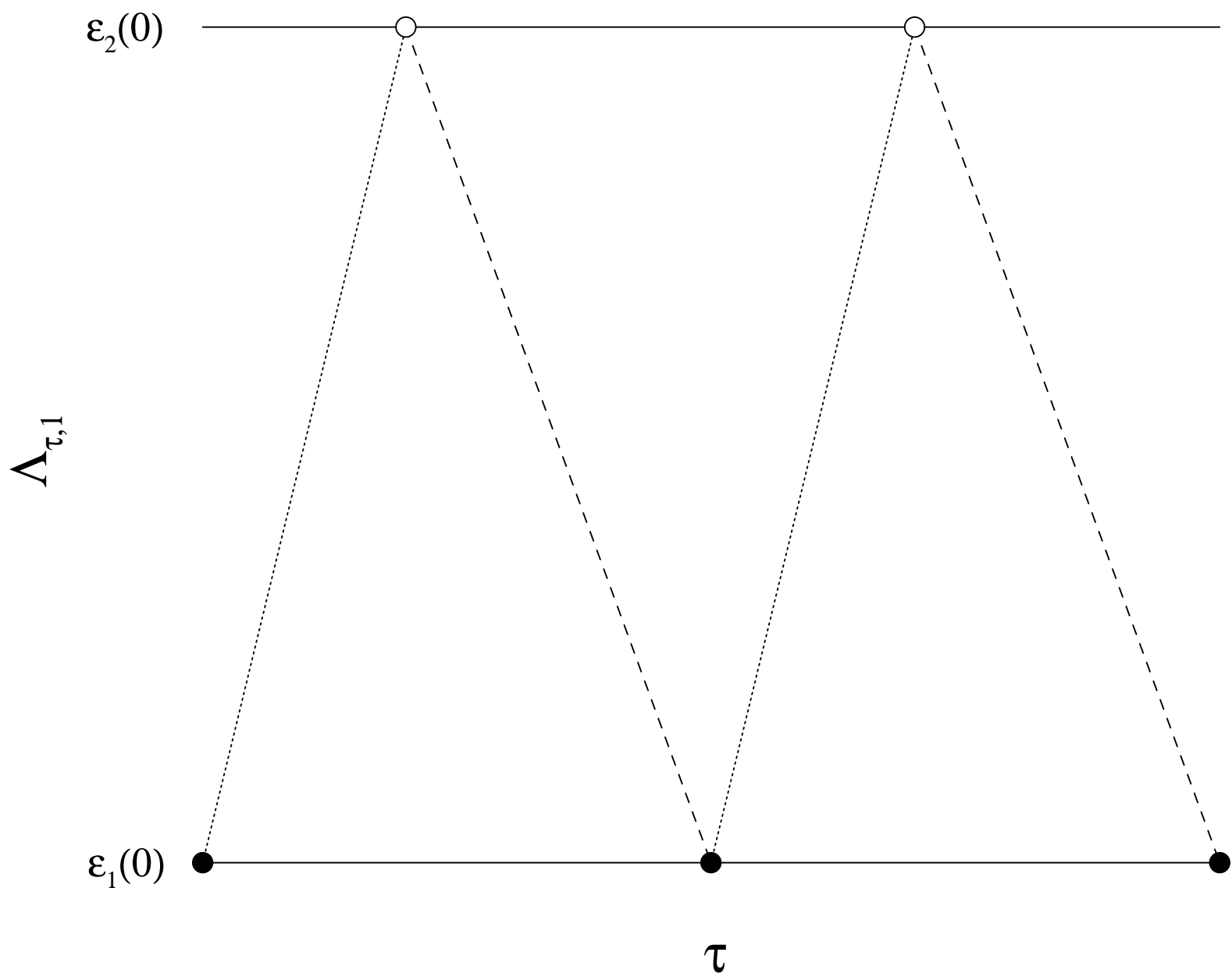
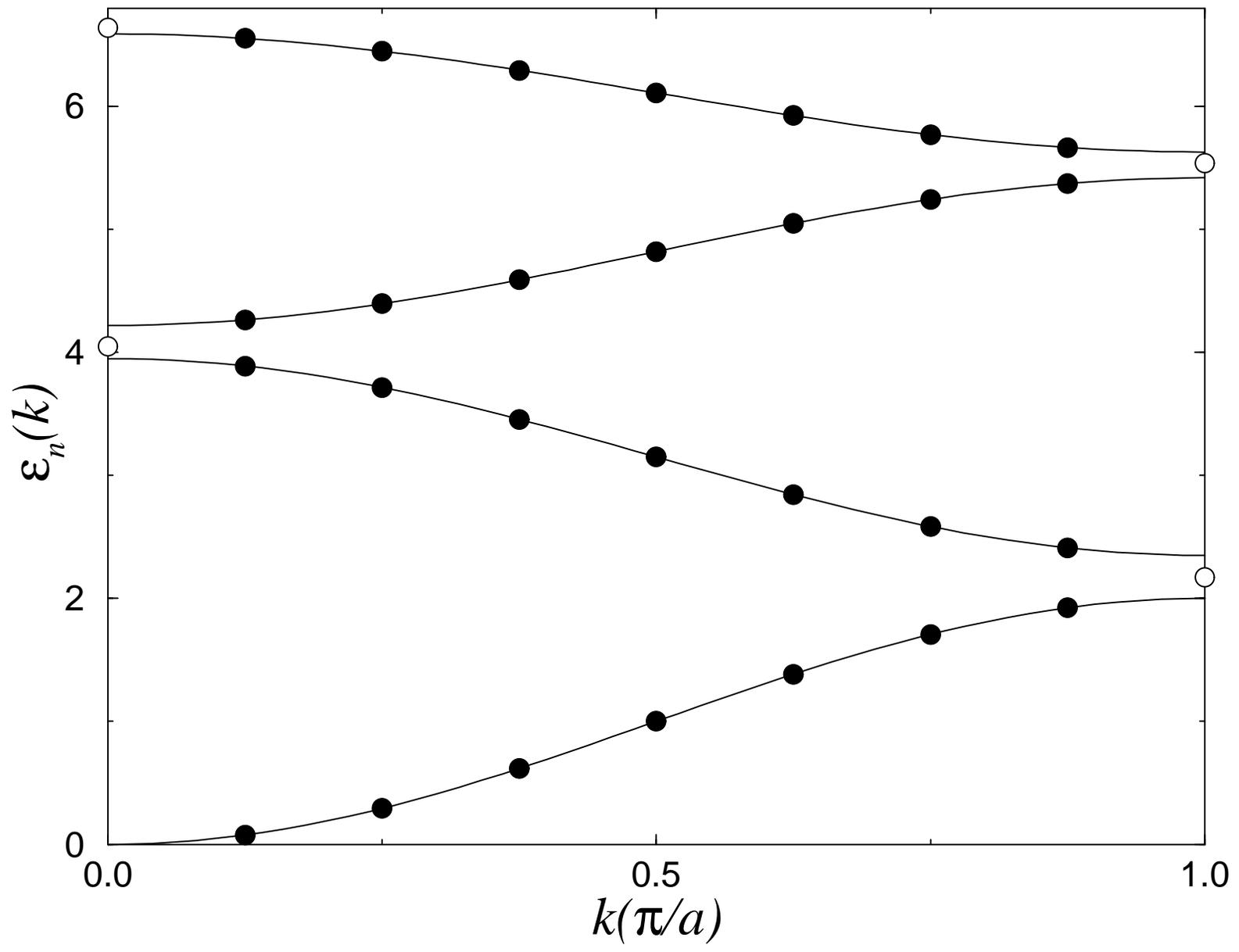
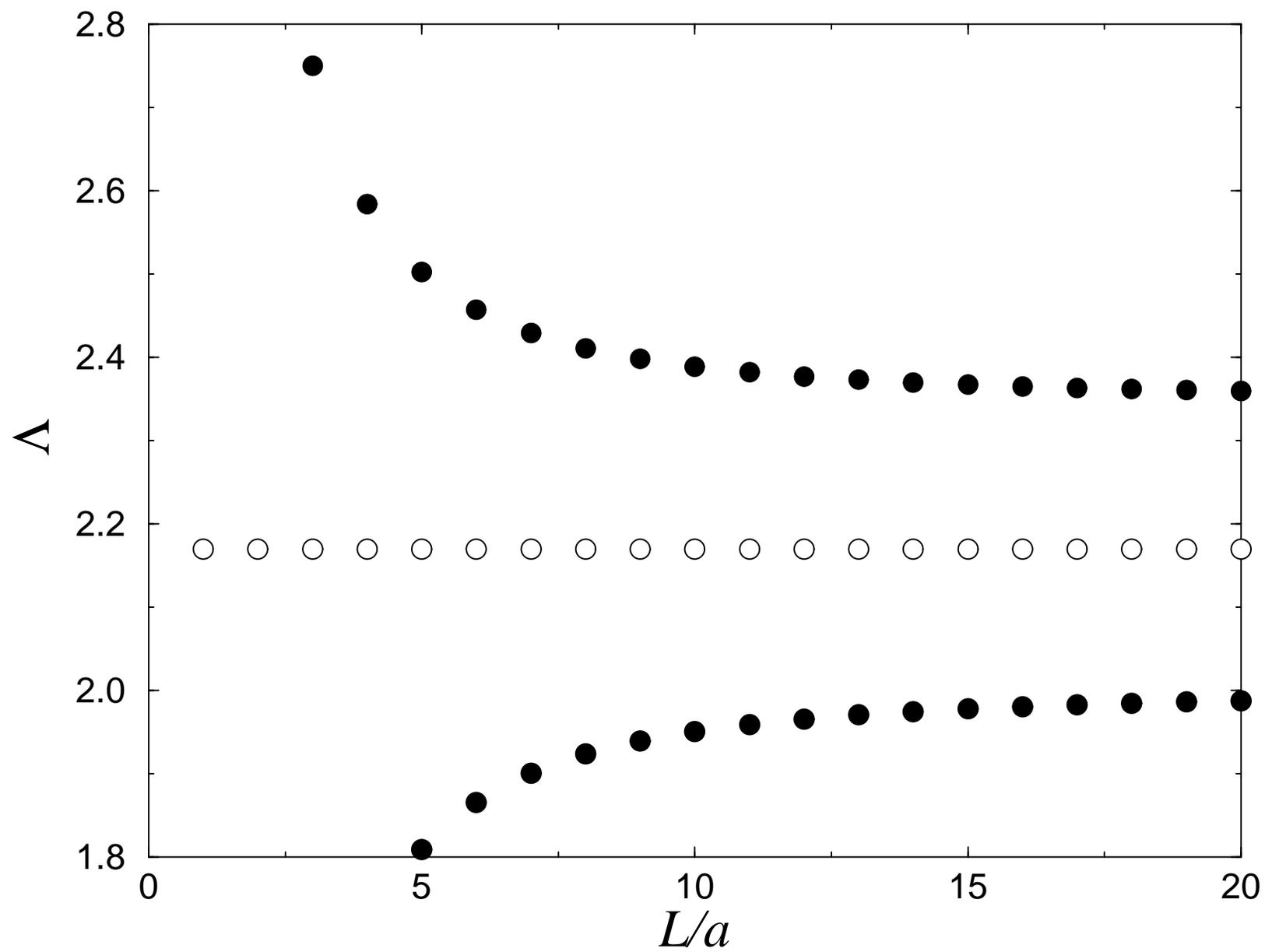
\end{document}